# Optimal Power Flow with Disjoint Prohibited Zones: New Formulation and Solutions

Xian Liu
Systems Engineering Department
Universiy of Arkansas at Little Rock
Little Rock, AR, USA

Hamzeh Davarikia
Electrical and Computer Engineering Department
Louisiana State University
Baton Rouge, LA, USA

*Abstract*— The constraints induced by prohibited zones (PZs) were traditionally formulated as multiple disjoint regions. It was difficult to solve the optimal power flow (OPF) problems subject to the disjoint constraints. This paper proposes a new formulation for the OPF problem with PZs. The proposed formulation significantly expedites the algorithm implementation. The effectiveness of the new approach is verified by different methods including traditional optimization methods, PSO and particle swarm optimization with adaptive parameter control which is conducted on the IEEE 30-bus test system.

## NOTATIONS

For convenience, the main notations to be used are listed below:

$B_{ik}, G_{ik}$ : Elements of the bus admittance matrix
$|V_i|$ : Voltage magnitude at bus $i$
$f_i$ : Objective function of generator $i$
$m_i$ : Number of feasible zones of generator $i$
$n_b$ : Number of total buses
$n_g$ : Number of generators
$P_{di}$ : Real power load at bus $i$
$P_{gi}$ : Real power generation at bus $i$
$Q_{di}$ : Reactive power load at bus $i$
$Q_{gi}$ : Reactive power generation at bus $i$
$\theta_i$ : The phase angle at bus $i$

## I. INTRODUCTION

The prohibited zone (PZ) of generators is a restricted operation range in which undesirable effects, such as the shaft bearing vibration, may get amplified [1]. While there exists a wealth of literature that deals with detection [2] and damping the oscillation in power grid's generators [3], prevention of happening this phenomenon is quite essential for power system reliability and resilience. The power grid is always faced with the risks, including the physical attack and the natural hazards [4] and the threats imposed from the nature of the power networks like the dynamic instability [5]-[7]. Various cyber-based approaches like real-time monitoring of the system [8] and data security enhancement [9], [10] have been proposed to improve the system resilience and security, but there is also a need to concentrate on physical-based approaches. While the optimization approaches are widely used in power system operation, resilience and reliability improvement, the PZ of generators are rarely used in the studies mentioned above due to the difficulties that introduced by PZs [11]-[20].

The optimal power flow (OPF) problem, with disjoint PZs as constraints, can be formulated as follows [21]:

(OPF_PZ0)

$$\text{minimize } y = \sum_{i=1}^{n_g} f_i(P_{gi}); \quad (1)$$

subject to

$$P_{gi} - P_{di} = |V_i| \sum_{k=1}^{n_b} |V_k| [G_{ik} \cos(\theta_i - \theta_k) + B_{ik} \sin(\theta_i - \theta_k)],$$

$$Q_{gi} - Q_{di} = |V_i| \sum_{k=1}^{n_b} |V_k| [G_{ik} \sin(\theta_i - \theta_k) - B_{ik} \cos(\theta_i - \theta_k)],$$

$$P_{i,\min} = a_{i1} \leq P_{gi} \leq b_{i1}, \text{ or } a_{i2} \leq P_{gi} \leq b_{i2}, \text{ or } \cdots,$$
$$\text{or } a_{im_i} \leq P_{gi} \leq b_{im_i} = P_{i,\max}, \quad (2)$$

$$Q_{i,\min} \leq Q_{gi} \leq Q_{i,\max}, \ (i=1,2,\cdots,n_g) \quad (3)$$

$$|V_k|_{\min} \leq V_k \leq |V_k|_{\max}. \ (k=1,2,\cdots,n_b) \quad (4)$$

In the above formulation,

$$a_{i1} < b_{i1} < a_{i2} < \cdots < a_{im_i} < b_{im_i}, (i=1,2,\cdots,n_g) \quad (5)$$

where $m_i$ is the number of feasible zones (FZs) of the generator with index "$i$", which implies that the number of PZs is $m_i$-1 Moreover, the reactive power is also subject to the constraints induced by (2), as the reactive power is an derived entity for the P-V buses and the slack bus. Note that the distinct feature of (2) is the relation "or". Namely, the generator $i$ is allowed to generate real power in any one of $m_i$ FZs. Accordingly, for the system of $n_g$ generators, the concerned OPF model consists of $m_r$ disjoint FZs, where $M_r = \sum_{i=1}^{n_g} m_i$, and $N_r$ sub-problems, where $N_r = \prod_{i=1}^{n_g} m_i$, due to the "or" relations. In principle, the OPF problem with (2) could be solved, in a brute-force manner, by solving $N_r$ sub-problems one by one, each with a contiguous FZ. Then the best solution of these sub-problems could be used as the solution to the original problem. However, this kind of approach in nature is an exhaustive enumeration procedure and becomes very inefficient when $N_r$ is large. In the traditional approaches, some sorts of programming commands or MIQP [22] method are used to deal with the PZs. This paper introduces a novel scheme to convert the disjoint FZs, expressed by (2) in OPF_PZ0 above, to an equivalent formulation expressed by a product of multiple terms. The resulting new model is solved by four methods; (a) common continuous optimization program (FMINCON built-in function in MATLAB), (b) conventional optimization method incorporating the *full load average production cost* (FLAPC) procedure, (c) PSO and (d) adaptive parameter control PSO (APSO) [23],[24].

The rest of this paper is organized as follows. In Section II, the new formulation of the OPF problem with disjoint PZs is presented. Then, in Section III, PSO with adaptive parameter control is discussed. The new model is tested on the IEEE 30-bus test system with four different methods in section IV. Finally, the conclusion is included in Section V.

## II. NEW FORMULATION

Solving the problem with disjoint working zones has always been a challenging issue, since this type of constraints is not algebraic, as shown in (2), and was conventionally implemented by series of programming commands. The conventional schemes, however, increased the programming complexity and were prone to divergence when searching for the optimal solution in some large-scale problems. Recently, a new algebraic method was presented for prohibited zones formulation [25], which may improve solving efficiency in terms of time and iteration. According to [25], the constraint set (2) in the model OPF_PZ0 is equivalent to the following form:

$$g(P_{gi}) = \prod_{k=1}^{m_i}(P_{gi}-a_{ik})(P_{gi}-b_{ik}) \leq 0. \quad (i=1,2,\cdots,n_g) \quad (6)$$

Accordingly, we have:

(OPF_PZ1)

$$\text{minimize} \quad y = \sum_{i=1}^{n_g} f_i(P_{gi}); \quad (7)$$

subject to

$$P_{gi} - P_{di} = |V_i|\sum_{k=1}^{n_b}|V_k|[G_{ik}\cos(\theta_i-\theta_k)+B_{ik}\sin(\theta_i-\theta_k)],$$

$$Q_{gi} - Q_{di} = |V_i|\sum_{k=1}^{n_b}|V_k|[G_{ik}\sin(\theta_i-\theta_k)-B_{ik}\cos(\theta_i-\theta_k)],$$

$$g(P_{gi}) = \prod_{k=1}^{m_i}(P_{gi}-a_{ik})(P_{gi}-b_{ik}) \leq 0, \quad (8)$$

$$Q_{i,\min} \leq Q_{gi} \leq Q_{i,\max}, \quad (i=1,2,\cdots,n_g) \quad (9)$$

$$|V_k|_{\min} \leq V_k \leq |V_k|_{\max}. \quad (k=1,2,\cdots,n_b) \quad (10)$$

## III. PSO WITH ADAPTIVE PARAMETER CONTROL

PSO was originally introduced by Kennedy and Eberhart in 1995 [26]. It is an evolutionary computational method in which the population is initialized randomly in the feasible range called particles. Each individual (i.e. particle) flies in the multidimensional search space to approach the optima, where each point represents a solution to the problem. For each particle there are two vectors, position, and velocity, which are modified according to the following equations [26]:

$$v_i(t+1) = wv_i(t) + c_1.r_1(pbest_i - x_i(t)) + c_2.r_2(gbest - x_i(t))$$

$$x_i(t+1) = x_i(t+1) + v_i(t+1) \quad (11)$$

where $pbest_i$ is the best location found by the $i^{th}$ particle, $gbest$ is the best location found by the entire swarm, $x_i$ is position and $v_i$ is the velocity of the $i^{th}$ particle and three control parameters are inertia weight $w$, cognitive learning factor $c_1$ and social learning factor $c_2$. Although PSO is a powerful technique for solving an optimization problem, its performance is highly affected by three control parameters. to ensure that the global optimum is reached especially in high non-convex problem like as OPF, these parameters should be tuned precisely which needs several trial-and-error for each problem. In [23] an adaptive approach for controlling the parameters during the optimization process has been proposed. In this method, $w$ is updated during the process as follows:

$$w = w_1.\exp(-gen/Maxgen) \quad (12)$$
$$w_1 = w_{high} - (w_{high} - w_{low}) \times (gen/Maxgen)$$

where $gen$ is a number of generation and $Maxgen$ is the number of maximum allowed generations. In addition, $w_{high}$ and $w_{low}$ are predefined boundary values. As in [23], $w_{high}$ and $w_{low}$ are set to 0.9 and 0.4, respectively. Other parameters of PSO are the learning factors $c_1$ and $c_2$ which decide the local and global search ability. According to (11), larger $c_1$ causes the current particle is highly affected by its previous best particle. On the other hand, when $c_2$ becomes larger, the current particle is more influenced by the global best particle. To dynamically adjust these factors [23] and [24] proposed a self-adaptive mechanism where $c_1$ and $c_2$ are as follows:

$$c_{1i} = c_{high} - (c_{high} - c_{low}) \times score_i \quad (13)$$

$$score_i = \begin{cases} \dfrac{f_{worst} - f(pbest_i)}{f_{worst} - f_{best}} & \text{if } f_{worst} > f_{best} \\ 1 & \text{otherwise} \end{cases}$$

where $f_{worst}$ and $f_{best}$ are the worst and best fitness values of the particle's best location respectively, and $f(pbest_i)$ is the best fitness value found by the ith particle. For better result, $c_{high}$ should be decreased linearly to $c_{low}$ during the process. The selected values of $c_{high}$ and $c_{low}$ are 2.5 and 0.5, respectively.

## IV. NUMERICAL EXPERIMENTS

As a verification for OPF_PZ effectiveness, the following four approaches, (a) conventional optimization method without FLAPC, (b) conventional optimization method with FLAPC [27], (c) conventional PSO, and (d) APSO, have been applied for solving the OPF problem on the IEEE 30-bus test system which consists of 41 lines and 6 generators. The bus-indexing specified originally in [28] is adopted in the present work. Accordingly, the bus attached by the bus with index "1" is chosen as the slack bus, with $V_1 = 1.06\angle 0$. The total load is 283.4 (MW). The mathematical formulation of OPF_PZ can be considered as an optimization problem as follow:

$$\begin{aligned} \text{minimize} \quad & f(x) \\ \text{subject to} \quad & g(x,u) = 0 \\ & h(x,u) \leq 0 \end{aligned} \quad (14)$$

where $x$ is a set of control variables including $P_{gi}$ and $u$ is a set of dependent variables, which includes $Q_{gi}$, $P_{di}$, $Q_{di}$, $|V_k|$, $\theta_i$. In the experiment, the cost metric is used in the objective function as (15):

$$f_i(P_i) = \alpha_i P_{gi}^2 + \beta_i P_{gi} + \gamma_i. \quad (15)$$

In the present study, the model OPF_PZ is solved with two layers. The master layer conducts the iterations for decision variables (real power), calls a subroutine that solves the power flow (PF) problem with equality constraints $G(x,u)$, calculates the objective function, and handles the inequality constraints $h(x,u)$. The PF subroutine is put into the second layer. The parameters used for performing PZs OPF in the test bed system are presented in Tables I through III. For the P-Q buses, $|V_i|_{min} = 0.95$ (p.u.) and $|V_i|_{max} = 1.07$ (p.u.).

TABLE I.  GENERATOR PARAMETERS

| $i$ | $P_{i,min}$ (MW) | $P_{i,max}$ (MW) | $Q_{i,min}$ (MVA) | $Q_{i,max}$ (MVA) | $|V_i|_{min}$ (p.u.) | $|V_i|_{max}$ (p.u.) |
|---|---|---|---|---|---|---|
| 1 | 2 | 50 | --- | --- | 1.06 | 1.06 |
| 2 | 3 | 60 | -40 | 50 | 0.95 | 1.1 |
| 5 | 5 | 100 | -40 | 40 | 0.95 | 1.1 |
| 8 | 6 | 120 | -10 | 40 | 0.95 | 1.1 |
| 11 | 5 | 100 | -6 | 24 | 0.95 | 1.1 |
| 13 | 3 | 60 | -6 | 24 | 0.95 | 1.1 |

TABLE II.  COST COEFFICIENTS

| $i$ | $\alpha_i$ ($/MW²h) | $\beta_i$ ($/MWh) | $\gamma_i$ ($/h) |
|---|---|---|---|
| 1 | 0.01 | 2 | 10 |
| 2 | 0.012 | 1.5 | 10 |
| 5 | 0.004 | 1.8 | 20 |
| 8 | 0.006 | 1 | 10 |
| 11 | 0.004 | 1.8 | 20 |
| 13 | 0.01 | 1.5 | 10 |

TABLE III.  PROHIBITED ZONES

| Unit | Prohibited Zones (MW) |
|---|---|
| 1 | None |
| 2 | [15, 20], [30, 40] |
| 5 | [15, 20], [60, 70] |
| 8 | [15, 20], [70, 80] |
| 11 | [15, 20], [60, 70] |
| 13 | [15, 20], [30, 40] |

It is well known that, even with the contiguous FZs, the OPF problem is a non-convex problem. The introduction of multiple disjoint PZs worsens the situation. An example profile is illustrated in Fig. 1 to show the effects of two disjoint PZs.

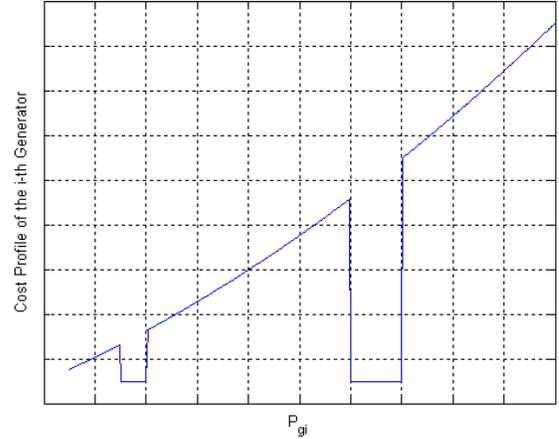

Fig. 1. An example of cost profile influenced by disjoint PZs.

The following methods have been applied to solve this high non-convex problem (14).

### A. Conventional optimization method

Built-in MATLAB function FMINCON has been applied to solve problem (14). In this process, Initial value $P_{0i}$ is selected randomly between [$P_{i,min}$, $P_{i,max}$].

### B. Conventional optimization method considering cfull load average production cost (FLAPC)

In this approach, (14) is solved by FMINCON considering FLAPC. Due to the problem's high non-convexity, how to select the starting points for the OPF program is an important issue. The adopted approach is based on the scheme of priority list. First, FLAPC for generator $i$ is defined as (16):

$$c_{ap,i} = \frac{f_i(P_{i,\max})}{P_{i,\max}}. \quad (\$/MWh) \quad (16)$$

Corresponding to the cost coefficients given in Table II, the FLAPC values for P-V buses are determined as shown in Table IV.

TABLE IV.  FLAPC VALUES

| $i$ | 2 | 5 | 8 | 11 | 13 |
|---|---|---|---|---|---|
| $c_{ap,i}$ | 2.3867 | 2.4000 | 1.8033 | 2.4000 | 2.2667 |

Then, a list called the priority list is made in the ascent order of the values of $c_{ap,i}$ as shown in Table VI, where the smaller numerals represent the higher priority. For example, when choosing the start point, the highest priority is given to the generator 8, followed by the generator 13, etc.

TABLE V. PRIORITY LIST

| i | 2 | 5 | 8 | 11 | 13 |
|---|---|---|---|----|----|
| Priority | 3 | 4 | 1 | 4 | 2 |

## C. Conventional particle swarm optimization (PSO)

In this approach, PSO has been applied to solve (14). For this issue, the constrained problem is transformed to an unconstrained one, by penalizing the constraints and building a new objective function. The objective function described in (14) is modified as follows using penalty function

$$\text{Min } \tilde{F}(x,u) =$$

$$f(x) \times \left(1 + \sum_{i=1}^{N_{eq}} \alpha_i G_i(x,u)^2 + \sum_{j=1}^{N_{ueq}} \beta_j \max[0, H_j(x,u)^2]\right) \quad (17)$$

where $\alpha_i$ and $\beta_j$ are penalty factors and $\tilde{F}$ is the objective function adopted for evaluating the fitness of each particle in population. In the current case study $\beta_j$ for the inequality constraint (8) and (10) is assumed 100 and $10^{-6}$ respectively. In addition, PSO parameters are shown in the Table VI.

TABLE VI. PSO PARAMETERS

| parameter | w | $c_1$ | $c_2$ | Swarm size | Max iteration |
|---|---|---|---|---|---|
| value | 0.95 | 2 | 2 | 100 | 50 |

## D. Adaptive control parameter PSO (APSO)

Proposed methods in III are applied to solve the OPF problem. Fig. 2 shows the procedure of solving OPF problem with PSO and APSO. Besides, main PSO parameters are considered as Table VI.

The numerical experiments are conducted successfully. Most tests successfully converged to the solutions presented in Tables VII. As a reference, the corresponding solutions for the PZ-free case are listed in Tables VIII and the solution with traditional approach is presented in Table IX. In these tables, the units of $P_{gi}$ is MW. The effects of PZs and different methods for solving this high non-convex problem can be clearly observed in the following tables and Fig.3- Fig.6.

TABLE VII. SOLUTION WITH PZS (NEW FORMULATION)

| Method | FMINCON | FLAPC | PSO | APSO |
|---|---|---|---|---|
| $P_{g1}$ | 12.0992 | 11.6276 | 11.818 | 10.8498 |
| $P_{g2}$ | 30.9517 | 29.2219 | 29.2693 | 29.9909 |
| $P_{g5}$ | 60.8014 | 58.0513 | 57.7793 | 58.0433 |
| $P_{g8}$ | 99.3513 | 97.3408 | 97.7033 | 97.3392 |
| $P_{g11}$ | 53.002 | 50.0502 | 49.6906 | 50.0686 |
| $P_{g13}$ | 29.9999 | 40 | 40.0433 | 40 |
| $\sum_i P_{gi}$ | 286.2055 | 286.2918 | 286.3038 | 286.2918 |
| $y_{min}$ (\$/h) | 607.0316 | 606.9621 | 606.9716 | 606.9501 |

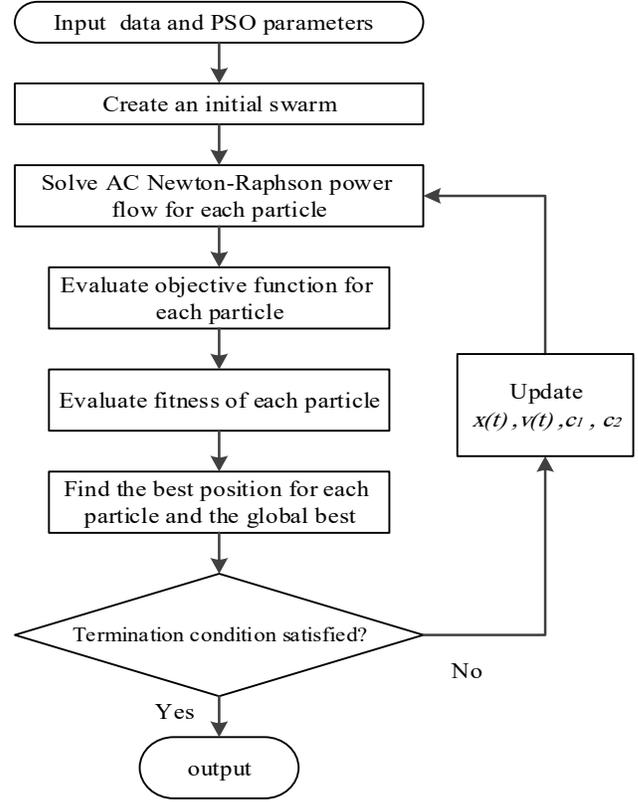

Fig. 2. APSO and PSO procedure

TABLE VIII. SOLUTION WITHOUT PZS

| Method | FMINCON | FLAPC | PSO | APSO |
|---|---|---|---|---|
| $P_{g1}$ | 11.436 | 11.4372 | 11.4599 | 11.4369 |
| $P_{g2}$ | 30.4027 | 30.4196 | 30.3525 | 30.4025 |
| $P_{g5}$ | 59.3481 | 59.3576 | 59.3682 | 59.3473 |
| $P_{g8}$ | 98.2555 | 98.26 | 98.2471 | 98.2559 |
| $P_{g11}$ | 51.4557 | 51.4357 | 51.4626 | 51.4559 |
| $P_{g13}$ | 35.3463 | 35.3338 | 35.3536 | 35.3459 |
| $\sum_i P_{gi}$ | 286.2443 | 286.2439 | 286.2439 | 286.2444 |
| $y_{min}$ (\$/h) | 606.6957 | 606.6957 | 606.6957 | 606.6957 |

TABLE IX. SOLUTION WITH PZS (TRADITIONAL APPROACH)

| Method | FMINCON | FLAPC | PSO | APSO |
|---|---|---|---|---|
| $P_{g1}$ | 25.875 | 14.4459 | 13.0456 | 13.2088 |
| $P_{g2}$ | 40 | 32.9633 | 32.1709 | 31.8937 |
| $P_{g5}$ | 70 | 60 | 63.4788 | 63.4451 |
| $P_{g8}$ | 20 | 80 | 80 | 80 |
| $P_{g11}$ | 70 | 60 | 60 | 60 |
| $P_{g13}$ | 60 | 38.56 | 37.1929 | 37.3424 |
| $\sum_i P_{gi}$ | 285.875 | 285.9692 | 285.8882 | 285.89 |
| $y_{min}$ ($/h) | 657.2453 | 609.3711 | 609.2713 | 609.2699 |

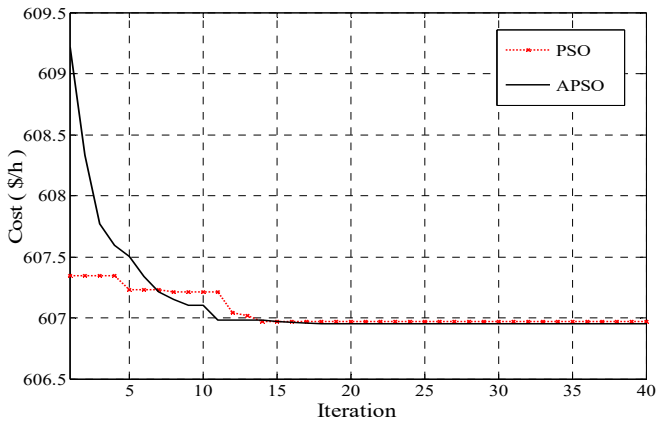

Fig. 3. Reducing cost during the PSO and APSO process in solving OPF_PZ

## V. CONCLUSION

The conventional approach for solving the OPF problem with the disjoint PZs usually relied on the enumeration schemes or random search schemes. These procedures become cumbersome when the number of such PZs increases. It is highly desirable to develop a new formulation to simplify the OPF model with PZs and expedite the implementation of the algorithms. The new approach presented in this paper represents such an effort. Numerical experiments on the IEEE 30-bus test system showed the effectiveness of the proposed approach. As the PZs increase non-convexity of OPF problem, traditional methods of optimization can hardly find global optima. It is shown that APSO has the best performance to find the global minimum of cost of generators. Besides, modification of the traditional approach which in FLAC and APSO indicates such reforms can notably improve the performance of original algorithm for solving more difficult problems.

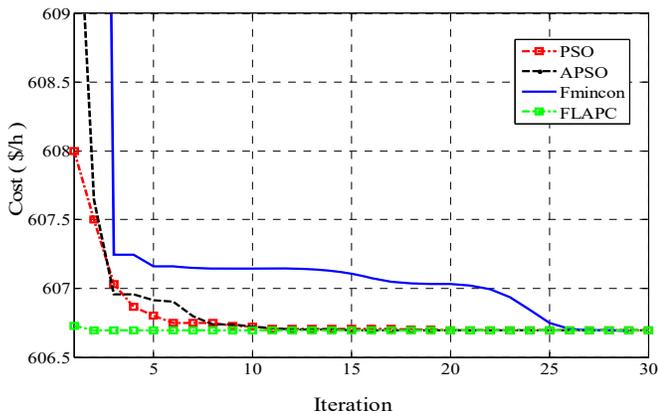

Fig. 4. Reducing cost during the PSO and APSO process in solving free prohibited zone OPF

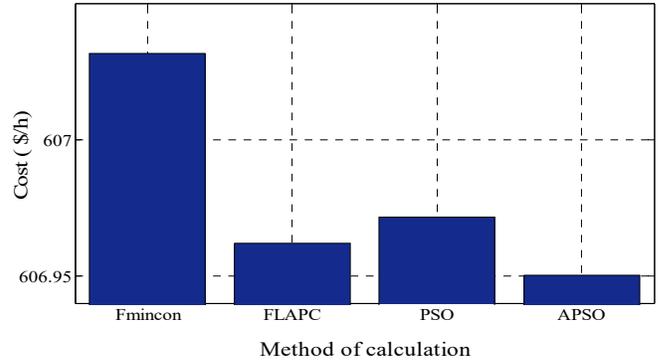

Fig. 5. Comparison between costs in four methods in solving OPF_PZ, new formulation

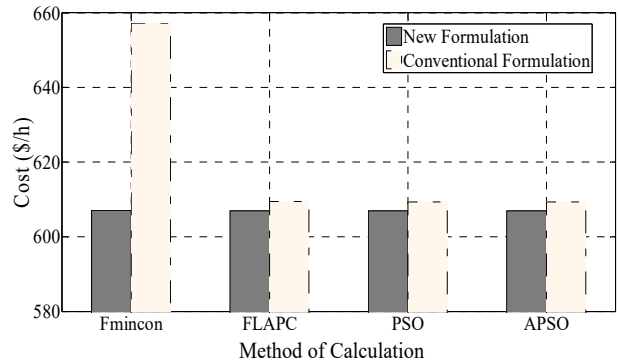

Fig. 6. Comparision between OPF_ PZ costs with new formulation and conventional approach